\documentclass[aps,floatfix,prd,showpacs]{revtex4}
\usepackage{graphicx}
\usepackage{dcolumn}
\usepackage{bm}

\begin{document}

\def\dq{\frac{d^4q}{(2\pi)^4}\,}
\def\dqE{\frac{d^4q_E}{(2\pi)^4}\,}
\def\be{\begin{equation}}
\def\ee{\end{equation}}

\title{The Gluon Propagator with Two-loop Schwinger-Dyson Equations}
\author{Joseph Meyers and Eric S. Swanson}
\affiliation{
Department of Physics and Astronomy, 
University of Pittsburgh, 
Pittsburgh, PA 15260, 
USA.}

\date{\today}

\begin{abstract}
Gluon, ghost, and quark propagators are computed in the Schwinger-Dyson formalism. The full set of one-loop and two-loop contributions to the gap equations are evaluated for the first time. A new and efficient method for dealing with quadratic divergences is presented and a consistent renormalization method is proposed. We find that the two-loop contribution to the propagators is subleading, good fits to lattice data are possible, and reliable vertex models are important to obtaining desired ultraviolet, infrared, and gauge-symmetric behavior.
\end{abstract}

\maketitle

\section{Introduction}

Developing analytic tools capable of describing nonperturbative phenomena of Quantum Chromodynamics (QCD) remains an elusive goal. In this regard, the gluon propagator is a nonperturbative quantity that is fundamental to QCD and has generated much interest. Early investigations with the Schwinger-Dyson formalism worked in Landau gauge and ignored the ghost contribution to the propagator. Although this approximation clearly violates fundamental properties of QCD, it was regarded as numerically reliable. Numerical and approximate solutions to the resulting gap equation yielded a propagator with a strong infrared enhancement, in keeping with simple expectations for QCD\cite{M}. 

Subsequently, it was realized that the ghost propagator can have an important influence on the Schwinger-Dyson equations and it was argued that the gluon and ghost propagators should behave as $(p^2)^{2\kappa}$ and $(p^2)^{-\kappa}$ respectively in the infrared limit ($\kappa$ is positive and of order unity)\cite{vSHA}. It was also noted that this expectation is in keeping with the confinement criterion of Kugo and Ojima\cite{KO}, which requires an unbroken realization of BRST symmetry (namely the Landau gauge ghost propagator should be more infrared singular than $(p^2)^{-1}$\cite{AvS}). 

A great deal of effort has gone into improving Schwinger-Dyson computations of the two-point functions in the intervening sixteen years. This effort has focussed on a variety of topics including improvements to vertex models\cite{sde-1}, inclusion of quarks and the derivation of chiral symmetry breaking\cite{sde-2,FWC}, the extension to finite temperature\cite{sde-3}, other gauges\cite{sde-4}, a background field implementation\cite{sde-5}, and investigations of various renormalization methods\cite{sde-6}. At the same time great improvements have occurred in lattice gauge methodology; this is of importance because it permits testing the effect of assumptions and truncations in the Schwinger-Dyson formalism.

The Schwinger-Dyson equations for the gluon are similar to those of abelian gauge theory with the addition of ghost and gluon loop contributions and two-loop contributions called the `squint' and `sunset' diagrams (see Fig. \ref{fig-gluon-gap}). To date the two-loop diagrams have not been fully considered, in large part because of difficulties in renormalizing overlapping divergences\cite{2loop}.
Although perturbation theory indicates that the two-loop diagrams are suppressed in the ultraviolet region, it is useful to confirm this in the infrared limit. Furthermore, until the two-loop contributions are obtained, it is impossible to disentangle the effects of vertex corrections from missing two-loop contributions.

We have therefore undertaken the computation of the full suite of diagrams contributing to the two-point functions of QCD in a variety of vertex truncations. Our results indicate that the two-loop terms do indeed make small contributions to the gluon propagator, with the squint diagram much larger than the sunset, and about 15\% of the size of the gluon loop diagram. 

The computation employs a Euclidean version of the momentum subtraction scheme for renormalization. Unfortunately this means that quadratic divergences appear that must be removed.  Current methods rely on performing a projection of the gluon polarization tensor to eliminate terms proportional to the metric or numerical procedures to isolate terms that contribute to quadratic divergences. Here we propose a different scheme based on standard renormalization methods that is simpler to implement. Finally, we argue that vertex models that are often employed in the literature are not natural and examine the effects of alternate vertices. Our results indicate that it is possible to get good agreement with lattice data and point the way to improvements in vertex  modelling.

\section{Gluon Gap Equations}

\subsection{Dressing Functions}

We seek to obtain nonperturbative expressions for the gluon, ghost, and quark propagators. The results will be defined in terms of dressing functions as follows:

\begin{eqnarray}
S(k) &\equiv& \int d^4x\, \langle T \psi(x)\bar \psi(0)\rangle{\rm e}^{ik\cdot x} = \frac{i}{A(k)\, \rlap{/}k - B(k)}, \\
\label{eq-D}
D_{\mu\nu}(k) &\equiv& \int d^4 x\, \langle T A_\mu(x) A_\nu(0)\rangle {\rm e}^{ik\cdot x} = -i\, P_{\mu\nu} \, G(k) - i \xi_0 k^\mu k^\nu F(k), \\
P_{\mu\nu}  &=& g_{\mu\nu} - \frac{k_\mu k_\nu}{k^2}, \\
H(k) &\equiv& \int d^4x\, \langle T c(x) \bar c(0)\rangle {\rm e}^{ik\cdot x} = i \frac{h(k)}{k^2}. 
\label{props}
\end{eqnarray}
Here $S$, $D_{\mu\nu}$, and $H$ are the full quark, gluon, and ghost propagators. Color indices have been suppressed in these expressions. Bare propagators are obtained by setting $A=1$, $B=m_0$, 
$G=1/k^2$, $F=1/k^4$, and $h=1$.  We shall also refer to variant dressing functions:

\begin{eqnarray}
M(k) &=& B(k)/A(k), \\
Z_q(k) &=& 1/A(k), \\
Z(k) &=& k^2 G(k).
\end{eqnarray}
Here $M$ is interpreted as the dynamical quark mass and $Z_q$ is the quark wavefunction. 

\begin{figure}[ht]
\includegraphics[width=4cm,angle=0]{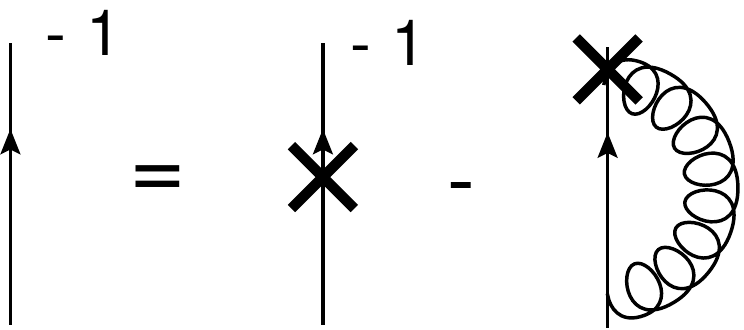}
\qquad\qquad
\qquad\qquad
\includegraphics[width=4cm,angle=0]{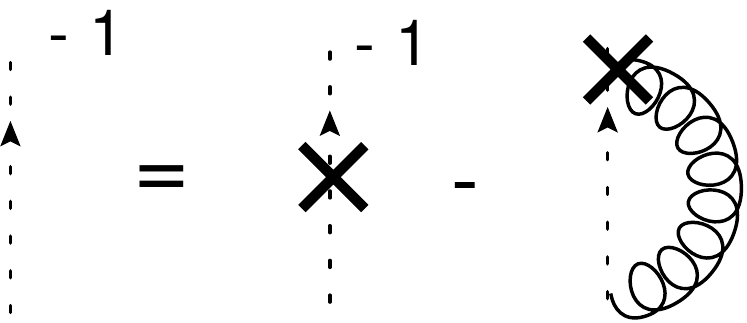}
\caption{Quark and Ghost Gap Equations. Crosses represent bare quantities.}
\label{fig-gap}
\end{figure}

\begin{figure}[ht]
\includegraphics[width=10cm,angle=0]{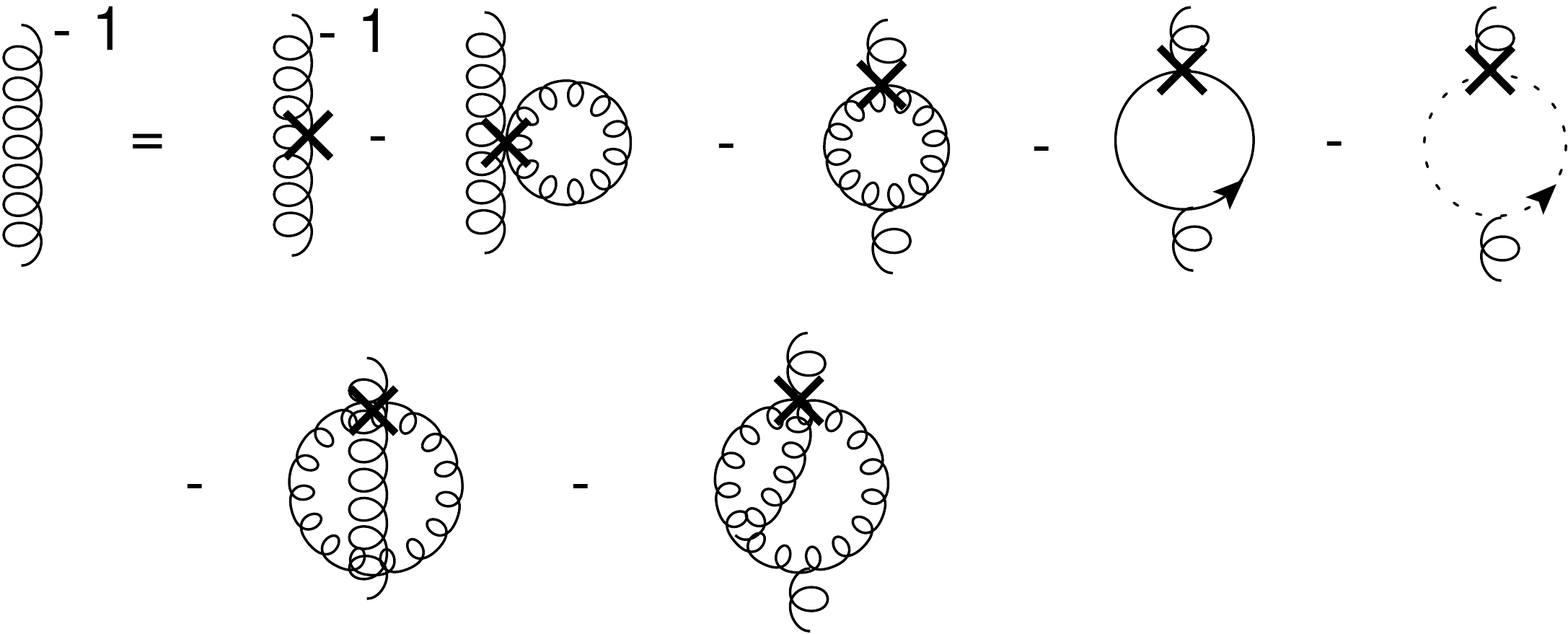}
\caption{Gluon Propagator Schwinger-Dyson Equation. The last two diagrams are the `sunset' and `squint' diagrams respectively.}
\label{fig-gluon-gap}
\end{figure}

The dressing functions are obtained by solving the Schwinger-Dyson equations shown in Figs. \ref{fig-gap} and \ref{fig-gluon-gap}. Each diagram has one bare vertex or propagator, which is denoted with a cross; all other quantities are dressed; external lines are stubs.

It will be convenient to refer to the gluon polarization tensor $\Pi_{\mu\nu}$. This can be defined in terms of the loop diagrams appearing on the right hand side of Fig. \ref{fig-gluon-gap} as

\be
(D)^{-1}_{\mu\nu}(k) = (D_0)^{-1}_{\mu\nu}(k) - i \Pi_{\mu\nu}(k).
\ee

\subsection{Model Vertices}
\label{vm}

The gap equations require quark-gluon, ghost-gluon, three-gluon, and four-gluon vertices. In principle these can be obtained from lattice computations or higher order Schwinger-Dyson equations. Although some progress in obtaining these vertices has been made\cite{sde-1}, the current state of the art is to model vertices in terms of known functions or functionals of the dressing functions, with the aim of closing the system of Schwinger-Dyson equations.

Care must be exercised in modelling vertices since too much freedom will simply lead to a parameterization of lattice (or other) data. Alternatively, imposing some expected behavior will often lead to vertices that are infinite as the cutoff is removed, and hence are not defined.

The simplest model uses bare vertices; this provides a useful starting point for the examination of nonperturbative phenomenology and can even be quantitatively accurate. For example, corrections to the ghost-gluon vertex are expected to be small\cite{ghost-gluon-vertex}.

In the case of abelian gauge theory, the Ward-Takahashi identity can be used to completely constrain the longitudinal dependence of the full fermion-gauge boson vertex. The result is given in terms of the fermion dressing functions, $A$ and $B$, and is called the Ball-Chiu vertex\cite{BC}. The situation is more complicated in the nonabelian case because the ghost dressing function and the quark-ghost scattering kernel, $K_g$ appear in the Slavnov-Taylor identity:

\be
k_\nu \Gamma_q^{a\, \nu}(p_1,p_2,k) = i g T^a h(k)[ S^{-1}(-p_1) K_g(p_1,p_2,k) - \bar{K_g}(p_2,p_1,k) S^{-1}(p_2)].
\ee

The resulting vertex has a rich structure that has been studied by Fischer and others\cite{quark-gluon-vertex}.
An adequate approximation to the full longitudinal structure of the vertex can be obtained by neglecting the quark-gluon scattering kernel and multiplying the Ball-Chiu vertex by the ghost dressing function. We make the further simplification of retaining only the `central' term of the Ball-Chiu vertex (called CBC below):
\be
\Gamma_q^{a\, \nu}(p_1,p_2,k) = - i g T^a \,\gamma^\nu\, h^n(p_1-p_2)\, \frac{1}{2}[A(p_1)+A(p_2)].
\label{qg-vertex}
\ee

A well-known shortcoming of the Ball-Chiu approach is that the transverse part of the vertex function remains unspecified. This has been partly remedied in the Abelian case by requiring multiplicative renormalizability in the asymptotic regime\cite{sde-6}. No such Ansatz exists for the nonabelian case and we choose to neglect this contribution henceforth. However, the correct perturbative asymptotic behavior for the quark propagator is not obtained with this approximation -- presumably due to the missing transverse components of the vertex. It has been observed that supplying an extra power of the ghost propagator in Eq. \ref{qg-vertex} ($n=2$) allows recovery of the expected asymptotic behavior\cite{quark-gluon-vertex}. We therefore choose to allow for varying values of $n$ in the following.

The ghost-gluon vertex has been considered by von Smekal {\it et al.}\cite{vSHA} who 
neglect the ghost-ghost scattering contribution to the vertex and retain only the reducible part of the four-ghost scattering matrix element. The result is an Ansatz of the form

\be
{\Gamma_\nu^{abc}}(q,p) =  g f^{abc}  \left[q_\nu \frac{h(k)}{h(q)} +  p_\nu \left( \frac{h(k)}{h(p)} -1\right)\right]
\label{eqn-H}
\ee
where $q$ is the incoming ghost momentum, $p$ is the outgoing ghost momentum, $k = p-q$ is the incoming gluon momentum. The ghost dressing is of order unity and slowly varying, so this vertex is indeed close to the bare vertex, as stated above. Finally, the bare four-gluon vertex is used in the following; the bare three-gluon vertex is also used except for a brief exploration with a simple model that is discussed in section \ref{results}.

%
%
%
%

\subsection{Quadratic Divergences}

The Schwinger-Dyson equations are renormalized by implementing a momentum subtraction scheme in Euclidean space. This amounts to performing subtractions in integral equations that substitute renormalization factors for constraints (the latter are analogous to observables such as pole location and propagator residue). Although this procedure is ideal for numerical problems it generates quadratic divergences that do not occur with other regulators such as dimensional regularization. A common method to deal with quadratic divergences to date is to project terms proportional to the metric tensor out of the vacuum polarization with the `Brown-Pennington' projector\cite{BP}. However this does not lead to satisfactory infrared behavior of the gluon dressing function\cite{FWC}, hence an alternative procedure that involves numerically isolating the quadratic divergence has also been used\cite{FWC,FAR}.

Here we propose a simpler method that is based on standard renormalization. Namely, a bare gluon mass term is included in the Lagrangian (the last term of Eq. \ref{L0}).



\begin{eqnarray}
{\cal L}_0 &=& -\frac{1}{4}Z_A (\partial_\mu A_\nu^a - \partial_\nu A_\mu^a)^2  + i Z_F \bar \psi \rlap{/}\partial \psi 
- Z_F m_0 \bar\psi \psi - Z_c {\bar c}^a \partial^2 c^a \nonumber \\
&&  + g_0 Z_F Z_A^{1/2} \bar \psi \rlap{/}A \psi - g_0 Z_A^{3/2} f^{abc}(\partial_\mu A_\nu^a) A_\mu^b A_\nu^c - g_0^2 Z_A^2 (f^{eab}A_\mu^a A_\nu^b)(f^{ecd} A_\mu^c A_\nu^d) \nonumber \\
&& -g_0 Z_c Z_A^{1/2} f^{abc}{\bar c}^a \partial_\mu A_\mu^b c^c - \frac{1}{2\xi_0} Z_A (\partial_\mu A_\mu^a)^2 + \frac{1}{2} Z_A \mu_0^2 (A_\mu^a)^2.
\label{L0}
\end{eqnarray}

This may look unusual since modern treatments typically have gauge invariant bare Lagrangians. However, this was a common procedure before the days of dimensional regularization\cite{JR}. The philosophical stance is that it is only the symmetry properties of the renormalized theory that are relevant. In this case the gluon mass term simply absorbs a quadratic divergence in the vacuum  polarization and hence gauge symmetry of the renormalized theory is guaranteed. With this approach the vacuum polarization is generalized to

\be
i\Pi_{\mu\nu} = i \Pi P_{\mu\nu} + i \Pi_1 \hat p_\mu \hat p_\nu
\ee
where $\Pi_1$ is a spurious form factor due to the regulator. This leads to a full gluon propagator:

\be
i D_{\mu\nu}(k) = \frac{-i P_{\mu\nu}}{Z_Ak^2 - Z_A \mu_0^2 -\Pi} - \frac{i \xi_0 \hat k_\mu \hat k_\nu}{Z_Ak^2 - Z_A\mu_0^2 \xi_0 - \xi_0 \Pi_1}.
\label{eq-DD}
\ee
The dressing function $F(k)$ of Eq. \ref{eq-D} involves the gauge non-invariant polarization scalar $\Pi_1$. However both functions become moot in Landau gauge, while $G$ is related to the standard vacuum polarization scalar 

\be
\Pi = \frac{1}{3} P^{\mu\nu}\Pi_{\mu\nu}.
\label{eq-pi}
\ee
In this way the usual structure of the gluonic propagator is recovered, with the exception of an additional renormalization factor that can be used to absorb the quadratic divergence in $\Pi$.

We stress that this method leads to different expressions than the Brown-Pennington method. Specifically, the usual BP approach simply uses $\Pi_{\rm BP} = \frac{1}{3} (g_{\mu\nu}- 4 k_\mu k_\nu/k^2) \Pi^{\mu\nu}$ in place of $\Pi$. This expression does not have a quadratic divergence, as desired, but it has a different, and incorrect, subleading structure. In fact, the presence of quadratic divergences forces the existence of a new form factor proportional to $g_{\mu\nu}$ and accounting for this changes the expression for the polarization scalar to agree with the conventional one of Eq. \ref{eq-pi}.

\subsection{Renormalization Scheme}

The right hand sides of the gap equations contain divergences which must be renormalized. We thus introduce multiplicative renormalization factors in the traditional way, as shown in Eq \ref{L0}.
As mentioned in the Introduction, it is most convenient to renormalize the gap equations using a momentum subtraction scheme in Euclidean space. A momentum space cutoff defined by a scale $\Lambda$ is introduced and a renormalization point denoted by $\mu$ (we sometimes consider different renormalization scales for different quantities). The gap equations  depend on the renormalization factors $Z_i(\mu,\Lambda)$. These are removed with renormalization conditions and yield renormalized vertices and renormalized dressing functions $A(k,\mu)$, $B(k,\mu)$, $h(k,\mu)$, and $G(k,\mu)$.

Care must be exercised in renormalizing the gap equations since it is possible to make inconsistent renormalization choices that render the equations meaningless. Furthermore, many vertex models that are in common use conflate modelling and renormalization, making it difficult to separate model-dependence from other assumptions. For example many practitioners choose to employ vertex models that depend on the cutoff,

\be
\Gamma^\nu_q = Z_F(\mu,\Lambda) \gamma^\nu .
\ee
This cutoff dependence must be absorbed elsewhere in the gap equations and obfuscates the structure of both the vertices and the propagators.

We propose to avoid these issues by working exclusively with renormalized quantities. As we shall see, this carries the price of making gauge noninvariance manifest. However, we argue that this is desirable since it reveals plainly the limitations induced by truncation, leads to consistent equations, and can be improved in a systematic (and hopefully controlled) fashion.

The procedure involves writing bare Schwinger-Dyson equations and extracting appropriate renormalization factors. This yields equations of the form

\begin{eqnarray}
S^{-1}(p) &=& Z_F (\rlap{/}p - m_0) - g_0 Z_F \sqrt{Z_A}\int S G \Gamma_q, \\
h^{-1}(p) &=& = Z_c - g_0 Z_c \sqrt{Z_A} \int h G \Gamma_{c}, \\
G^{-1}(p) &=& Z_A p^2 - Z_A \mu_0^2 - g_0 Z_F \sqrt{Z_A}  \int S S \Gamma_q - g_0 Z_c \sqrt{Z_A} \int hh \Gamma_{c}  - \nonumber \\ 
&& g_0 Z_A^{3/2}  \int GG \Gamma_{3g} - g_0^2 Z_A^2 \int GGG\Gamma_{4g} - g_0^2 Z_A^2 \int GGGG\Gamma_{3g} \Gamma_{3g},
\end{eqnarray}
where all dressing functions are renormalized and $\Gamma_q$, $\Gamma_{c}$, $\Gamma_{3g}$, and $\Gamma_{4g}$ refer to renormalized quark-gluon, ghost-gluon, three-gluon, and four-gluon vertices. All tensor structure and factors of $i$ have been suppressed.

Standard expressions relating renormalized and bare dressing functions have been used to obtain these equations. For example

\be
S(p,\mu) = \frac{1}{Z_F(\mu,\Lambda)} S_0(p,\Lambda)
\label{eq-S}
\ee
with
\be
S_0 = \frac{i}{A_0(p,\Lambda)\rlap{/}p - B_0(p,\Lambda)},
\ee
while the one particle irreducible quark-gluon vertex obeys
\be
\Gamma_q^\nu(s,t,u,\mu) = Z_F \sqrt{Z_A} \Gamma^\nu_{q\,0}(s,t,u,\Lambda).
\label{eq-G}
\ee

Thus the tree level quark-gluon vertex model is
\be
\Gamma^\nu_{\rm tree} = g_0 Z_F \sqrt{Z_A} \gamma^\nu
\ee
with similar expression for the other vertices. 

Vertex models (or solutions to Schwinger-Dyson equations) can be renormalized by fixing their strength at a kinematic point, $s_\star,t_\star,u_\star$: e.g.,

\be
\Gamma_q^\nu(s_\star,t_\star,u_\star) = g_q\gamma^\nu.
\ee
Thus 
\be
g_0 Z_F \sqrt{Z_A} = g_q.
\label{g1}
\ee
Doing the same for the other vertices gives
\be
g_0 Z_c \sqrt{Z_A} = g_{c},
\label{g2}
\ee
\be
g_0 Z_A^{3/2} = g_3,
\label{g3}
\ee
and
\be
g_0 Z_A = g_4.
\label{g4}
\ee

Using tree level model vertices yields gap equations

\begin{eqnarray}
S^{-1}(p) &=& Z_F(\rlap{/}p - m_0) - g_q^2 \int S G(p), \nonumber \\
h^{-1}(p) &=& Z_c - g_{c}^2 \int h G(p),  \nonumber\\
G^{-1}(p) &=& Z_A p^2 - Z_A \mu_0^2 - g_q^2 \int SS(p) - g_{c}^2 \int hh(p) - \nonumber \\&&  g_3^2\int GG(p) - g_4^4 \int GGG(p) - g_3^2 g_4^2 \int GGGG(p).
\label{eq-renorm}
\end{eqnarray}
Finally, the field renormalization factors are identified by imposing renormalization conditions. For example setting the ghost dressing function to be $h(\mu)$ at $p=\mu$ gives

\be
Z_c(\mu,\Lambda) = h^{-1}(\mu) + g_{c}^2\int^\Lambda h G(\mu).
\label{Zc}
\ee

Two subtractions are required to define the gluon propagator since the gluon loop integrals are quadratically divergent. A convenient choice for determining $Z_A\mu_0^2$ is at $p^2=0$  since the quadratic divergence is absorbed by $G(0$) and gauge invariance is retained.

In contrast to this, the truncations implicit in Eqs \ref{eq-renorm} necessarily break gauge invariance. This can be seen in the lack of consistency  Eqs. \ref{g1} -- \ref{g4}. For example, Eq. \ref{g3} implies that $Z_A(\mu,\Lambda)^{3/2} \sim g_3(\mu)/g_0(\Lambda)$ which is not compatible with Eq \ref{g4}.  More generally, the four renormalization constants obtained as in Eq. \ref{Zc} are inconsistent with Eqs. \ref{g1} -- \ref{g4}.
This incompatibility must be due to the vertex model truncations and cannot be resolved except by abandoning presumed gauge invariance. A simple way to achieve this is to employ different bare couplings 

\be
g_0 \to g_{q\, 0},\ g_{c\, 0},\ {\rm etc}.
\ee
%

Presumably the lack of gauge invariance will be reflected in the values of the renormalized gauge couplings. Furthermore, one expects that gauge couplings will converge to a common value as the accuracy of the Schwinger-Dyson truncation is improved. Unfortunately this is difficult to quantify as the $n$-point truncation scheme is not tied to a small parameter. Nevertheless, the premise of this approach to QCD is that the tree order Schwinger-Dyson equations provide a quantitatively useful approximation to two-point functions. This can be justified {\it a posteriori} if the predictions prove accurate or if full vertices 
are not heavily dressed. 

Of course the implication is that increasingly accurate representations of QCD $n$-point functions yield more accurate propagators. For example, the schematic equation for the quark-gluon vertex (in the bare quark-gluon vertex form) reads

\be
\Gamma_q = g_0 Z_F\sqrt{Z_A} \left( \gamma + \int SSG \Gamma_q\Gamma_q + \int SGG \Gamma_q \Gamma_{3g} + \int SG \Gamma_{qqgg}\right).
\ee
This can be renormalized by setting $\Gamma_q(s_\star,t_\star,u_\star) = g_q$ to obtain $g_0 Z_F\sqrt{Z_A}$. Thus the vertex is determined along with one constraint on the renormalization factors. Again, these equations cannot be consistent if a model quark-quark-gluon-gluon vertex is employed, and different bare and renormalized couplings must again be used. 
However, it seems likely that the numerical difference between the renormalized couplings will be smaller, and will continue to get smaller as one marches up the $n$-point truncation of the Schwinger-Dyson equations and obtains results that 
obey the
strictures of gauge symmetry to a greater degree. We will check this in the following by comparing tree level couplings to those obtained with the vertex models of Section \ref{vm}.

\subsection{The Renormalized Gap Equations}

With renormalization in place we are ready to specify the gap equations.
The quark gap equation is obtained by projecting the quark propagator onto its two components. We work in Landau gauge and perform the Wick rotation to Euclidean space. The result before subtraction is

\be
B(p_E^2) = Z_F m_0 + 3 g_q^2 C_F \int \dqE W\, \sigma_S(q_E^2)\, G(Q_E),
\label{eq-B-gap}
\ee

\be
A(p_E^2) = Z_F - g_q^2 C_F \int \dqE W\, \sigma_V(q_E^2) \, G_E(Q_E)\,\left(\frac{2 q_E^2(1-x^2)}{Q_E^2} - 3 \frac{q_E x }{p_E}\right).
\label{eq-A-gap}
\ee
The quantities under the integrals are $\sigma_S(q) = B(q)/[A^2(q) q^2 + B^2(q)]$ and $\sigma_V(q) = A(q)/[A^2(q) q^2 + B^2(q)]$.

The ghost gap equation is 

\be
\frac{1}{h(p_E^2)} = Z_c -  g_{c}^2 C_A\, \int \frac{d^4q_E}{(2\pi)^4}\, V \, G_E(Q_E)\, h(q_E^2) \frac{1-x^2}{Q_E^2}.
\label{eq-h-gap}
\ee
The Euclidean nature of these expressions has been emphasized with the subscript. We have also defined  $Q = p-q$, $C_F = (N_c^2-1)/2N_c$, and $C_A = N_c$. By convention $G_E(Q_E) = - G(Q \to Q_E)$, and $B(p_E) = B(p\to p_E)$. The factors appearing in these expressions are due to vertex models, if present, and are

\be
V = \frac{h(Q)}{h(q)} + \frac{h(Q)}{h(p)} - 1
\label{eq-V}
\ee
and 
\be
W = \frac{1}{2} h^n(Q) \Big(A(p) + A(q)\Big).
\ee
Rainbow-ladder results are recovered when $h\to 1$ and $A\to 1$ in these expressions.

The equation for the gluon dressing function depends on the vacuum polarization scalars as indicated in Eq. \ref{eq-DD}.  Explicit expressions in the rainbow-ladder approximation are

\begin{eqnarray}
\Pi_{\rm gluon}(p) &=& -\frac{2}{3}\,g_3^2N_c\,\int\frac{d^4q_E}{(2\pi)^4}G\big(q^2\big)G\big((q-p)^2\big) \nonumber \\
&& \times\left(\frac{3p^4+3q^4-6p^3qx-6pq^3x+p^2q^2\left(8+x^2\right)}{p^2+q^2-2pqx}\right)\left(1-x^2\right),
\end{eqnarray}

\begin{eqnarray}
\Pi_{\rm quark}(p) &=& -\frac{1}{6}\, g_q^2N_c C_F\,\int\frac{d^4q_E}{(2\pi)^4}\Big[3\sigma_S\big((q-p)^2\big)\sigma_S\big(q^2\big) \nonumber \\
&& +\Big(q^2\left(1+2x^2\right) - 3qpx\Big)\sigma_V\big((q-p)^2\big)\sigma_V\big(q^2\big)\Big],
\end{eqnarray}

\be
\Pi_{\rm ghost}(p) = \frac{1}{3}\,g_{c}^2N_c\,\int\frac{d^4q_E}{(2\pi)^4}h\big(q^2\big)h\big((q-p)^2\big)\frac{1-x^2}{q^2+p^2-2qpx},
\ee

\begin{eqnarray}
 \Pi_\mathrm{squint}(p) &=& g_3^2 g_4^2 N_c^2\,\int\frac{d^4q_E\ d^4k_E}{(2\pi)^8}G\big(q^2\big)G\big(k^2\big)G\big((q+k)^2\big)G\big((p-q-k)^2\big) \nonumber \\
&& \times\bigg[q^6\big(3-3x^2-w^2(3-x^2)+2wxy\big) \nonumber \\
&& +k^6\big(3-3y^2-w^2(3-y^2)+2wxy\big) \nonumber \\
&& + q^5k\big(13w(1-w^2)-3wx^2(3-w^2)-2xy(3-5w^2)+2wy^2\big) \nonumber \\
&& +qk^5\big(13w(1-w^2)-3wy^2(3-w^2)-2xy(3-5w^2)+2wx^2\big) \nonumber \\
&& +q^4k^2\big(9-2x^2(3-w^4)+w^2(7-16w^2) \nonumber \\
&& -2wxy(8-7w^2)-3y^2(1-3w^2)\big) \nonumber \\
&& +q^2k^4\big(9-2y^2(3-w^4)+w^2(7-16w^2) \nonumber \\
&& -2wxy(8-7w^2)-3x^2(1-3w^2)\big) \nonumber \\
&& +q^3k^3\big(2w(13-11w^2-2w^4)-4xy(3-w^4)-w(x^2+y^2)(7-11w^2)\big) \nonumber \\
&& -pq^5\big(x(5-7w^2)-x^3(5-w^2)+2wy+4wx^2y\big) \nonumber \\
&& -pk^5\big(y(5-7w^2)-y^3(5-w^2)+2wx+4wxy^2\big) \nonumber \\
&& +pq^4k\big(2wx(5-8w^2)-wx^3(6-w^2)+y(5+w^2) \nonumber \\
&& -x^2y(15-13w^2)+7wxy^2\big) \nonumber \\
&& +pqk^4\big(2wy(5-8w^2)-wy^3(6-w^2)+x(5+w^2) \nonumber \\
&& -xy^2(15-13w^2)+7wx^2y\big) \nonumber \\
&& +pq^3k^2\big(-2x(5-5w^2-2w^4)+x^3(5-7w^2)+15xy^2 \nonumber \\
&& -4wy(3-2w^2)+wx^2y(16-7w^2)-3wy^3-19w^2xy^2\big) \nonumber \\
&& +pq^2k^3\big(-2y(5-5w^2-2w^4)+y^3(5-7w^2)+15x^2y \nonumber \\
&& -4wx(3-2w^2)+wxy^2(16-7w^2)-3wx^3-19w^2x^2y\big) \nonumber \\
&& +p^2q^4\big(5-x^2(5-w^2)-5w^2+4wxy\big) \nonumber \\
&& +p^2k^4\big(5-y^2(5-w^2)-5w^2+4wxy\big) \nonumber \\
&& +p^2q^3k\big(12w(1-w^2)-wx^2(6-w^2)-2xy(5-6w^2)+3wy^2\big) \nonumber \\
&& +p^2qk^3\big(12w(1-w^2)-wy^2(6-w^2)-2xy(5-6w^2)+3wx^2\big) \nonumber \\
&& +p^2q^2k^2\big(10-2w^2(3+2w^2)-2wxy(5-3w^2)-(x^2+y^2)(5-7w^2)\big)\bigg] \nonumber \\
&& \times\left[\frac{1}{(k^2+q^2+2qkw)(p^2+k^2+q^2-2pky-2pqx+2qkw)}\right],
\end{eqnarray}
and

\begin{eqnarray}
\Pi_\mathrm{sunset}(p) &=&  -\frac{1}{6}\,g_4^4 N_c^2\,\int\frac{d^4q_E\ d^4k_E}{(2\pi)^8}G\big(q^2\big)G\big(k^2\big)G\big((p-q-k)^2\big) \nonumber \\
&& \times \bigg[q^2\big(11+x^2+w^2(1+x^2)+4wxy\big) \nonumber \\
&& +k^2\big(11+y^2+w^2(1+y^2)+4wxy\big) \nonumber \\
&& +qk\big(w(23+2x^2+2y^2+w^2)+xy(3+5w^2)\big) \nonumber \\
&& -pq\big(23x+x^3+3wy+5wx^2y+2w^2x+2xy^2\big) \nonumber \\
&& -pk\big(23y+y^3+3wx+5wxy^2+2w^2y+2x^2y\big) \nonumber \\
&& +p^2\big(11+x^2+y^2+x^2y^2+4wxy\big)\bigg] \nonumber \\
&& \times\left[\frac{1}{p^2+k^2+q^2-2kpy-2qpx+2qkw}\right].
\end{eqnarray}
The angular variables in these expressions are defined as $x = \hat p \cdot \hat q$, $y = \hat p \cdot \hat k$, and $w = \hat q \cdot \hat k$. 
Finally, 

\be
G^{-1}(p) = Z_A p^2 +  Z_A \mu_0^2  + \Pi_{\rm quark} + \Pi_{\rm ghost} + \Pi_{\rm gluon} + \Pi_{\rm squint} + \Pi_{\rm sunset}.
\ee

As stated above, the remaining renormalization factors are absorbed by subtracting at appropriate scales. Thus, for example,

\be
\frac{1}{h(p_E^2)} = \frac{1}{h(\mu^2)} -  g_{c}^2 C_A\, \int \frac{d^4q_E}{(2\pi)^4}\, \left[ V\,  G_E(Q_E)\, h(q_E^2) \frac{1-x^2}{Q_E^2} - {\rm same}(p=\mu)\right].
\label{eq-h-gapR}
\ee
The integral is finite and the cutoff can safely be removed in this expression.

\section{Numerical Results and Comparison to the Lattice}
\label{results}

The coupled gap equations were solved numerically using Gauss-Seidel relaxation, modified Levenberg-Marquardt minimization, and iteration. We found that iteration worked very well over a wide variety of renormalization and coupling choices and this method was used to generate the majority of our results. Integrals were evaluated on a momentum grid. Typical grids had 48 Gauss-Legendre points in the radial direction and 24 in the angular direction. Runs with up to 98 radial and 48 angular points were not uncommon. We found that all loop integrals converged rapidly except the squint diagram, which required extrapolation.

\begin{figure}[ht]
\includegraphics[width=10cm,angle=0]{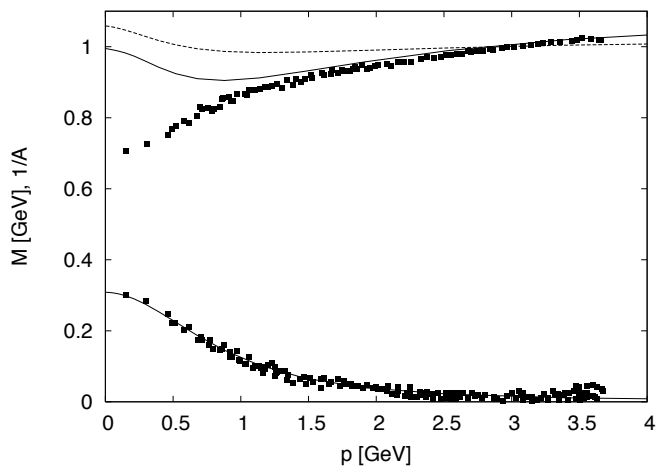}
\caption{The quark wavefunction renormalisation, $Z_q=1/A$, and dynamical quark mass $M$, along with lattice results\protect\cite{lattice-quark}.} 
\label{fig-quark-lattice}
\end{figure}

\begin{figure}[ht]
\includegraphics[width=10cm,angle=0]{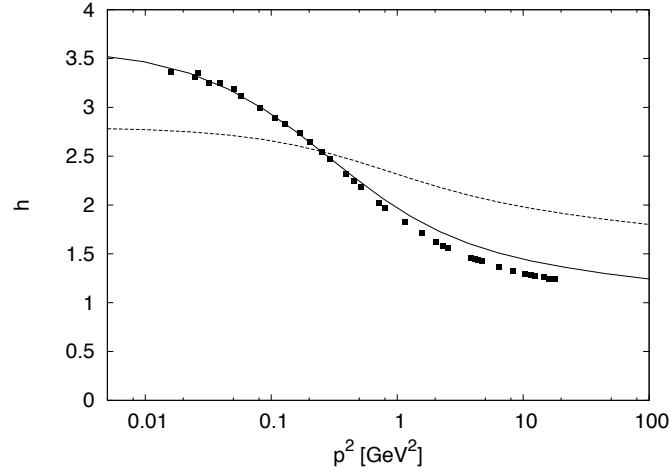}
\caption{Ghost Dressing Function and Lattice Data\protect\cite{lattice-gluon}.}
\label{fig-ghost-lattice}
\end{figure}

\begin{figure}[ht]
\includegraphics[width=10cm,angle=0]{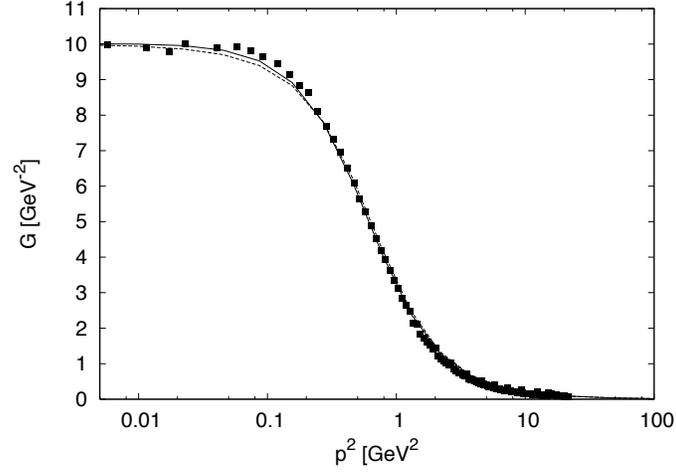}
\caption{Gluon Dressing Function and Lattice Data\protect\cite{lattice-gluon}.}
\label{fig-gluon-lattice}
\end{figure}

Renormalization was carried out by comparing to lattice computations of the gluon\cite{lattice-gluon, B2} and quark\cite{lattice-quark} propagators. 
Typical renormalization values used in the chiral case were $G(0) = 10$ GeV$^{-2}$, $G(0.5\  {\rm GeV}) = 8.0$ GeV$^{-2}$, $h(0.5\  {\rm GeV}) = 2.55$, and $A(3.0\  {\rm GeV}) = 1.0$. The latter value was chosen to agree with the renormalization condition used by Bowman {\it et al.}\cite{lattice-quark}. As a first task, we attempted a fit to the massless quark lattice data with a single coupling. This fit was made with tree level vertices (the rainbow-ladder approximation) and yielded $g \approx 0.6$. The results of this fit are displayed as dashed lines in Figs. \ref{fig-quark-lattice} - \ref{fig-gluon-lattice}. The figures make it clear that the magnitudes of the quark and ghost dressing functions are too small when a universal coupling is used.  In fact, there is not sufficient infrared strength in the quark gap equation to generate spontaneous chiral symmetry breaking (this has been noted before\cite{quark-gluon-vertex}).

We then attempted a fit allowing $g_{c}$, $g_q$, and $g_3= g_4$ to float with tree-level vertices. One obtains $g_3 \approx 0.6$ as before, however reasonable fits to the dynamical quark mass and the ghost dressing function require larger couplings, $g_{c} = 1.12$ and $g_q \approx 3.03$, indicating that substantial strength is missing in the tree ghost-gluon and quark-gluon vertices. This fit yielded a dynamical quark mass that scaled as $p^{-3/2}$ in the ultraviolet region, which is not the expected $\langle \bar \psi \psi\rangle/p^2$. Furthermore, the iterative procedure did not converge well and it was in fact difficult to tell if a stable solution exists, which may in part explain the unusual momentum scaling.

Employing the CBC vertex changed this behavior entirely, iteration converged well and the asymptotic behavior of the dynamical quark mass (in the chiral case) was the expected $1/p^2$ to great accuracy. The new coupling constants were 

\be
g_q = 1.35 \qquad g_{c} = 1.12 \qquad q_3 = 0.6.
\label{cc}
\ee
As hoped, these couplings are much closer to a universal value than those associated with tree level vertices.

This success was not reproduced with the model ghost-gluon vertex of Eq. \ref{eqn-H}; it tended to cause instability in the iteration and did not lead to improvement in the fit. Perhaps this is not surprising since it has been suggested that this vertex is not substantially modified.  This can be tested further by examining the ultraviolet behavior of the dressing functions. In general one expects that perturbative anomalous dimensions will dictate this behavior. One thus has

\be
G(p) \to p^{-2} \log(p)^{\gamma}
\ee
with

\be
\gamma = \frac{3\xi N_c - 13 N_c + 4 n_f}{22 N_c - 4 n_f}.
\label{eq-g}
\ee
Similarly $h \to \log(p)^{\gamma_h}$, with
\be
\gamma_h = \frac{3\xi N_c - 9 N_c}{44 N_c - 8 n_f}.
\ee
We note that the lattice data agree quite well with these predictions. Numerically,  $\gamma_h(n_f=0) = -0.205$ and $\gamma_h(n_f=1) = -0.218$ in Landau gauge.
The analogous results from the  Schwinger-Dyson equations are -0.20 and -0.21 respectively. Thus it appears that the ghost dressing function is following expectations quite well. 
We conclude that the tree ghost-gluon vertex is adequate for this investigation, in keeping with previous observations\cite{ghost-gluon-vertex}.

In view of these observations our second model uses a tree ghost-gluon vertex and the central Ball-Chiu vertex with differentiated couplings given in Eq. \ref{cc}. The resulting dressing functions are shown as solid lines in Figs. \ref{fig-quark-lattice} - \ref{fig-gluon-lattice}. The dynamical mass is reproduced extremely well even though no renormalization constant is available in the chiral limit considered here. The quark wavefunction is not well reproduced. 
This is likely due to the simple quark-gluon vertex we used. Namely, missing transverse structure is known to affect $A$ more strongly than $B$, and we suspect that this is the cause of the problem\cite{RW}.  It is thus fortunate that $A$ lies near unity over its entire range, and that the Schwinger-Dyson results do as well, never deviating by more than approximately 30\% from the lattice computation.

The ghost dressing function is also quite close to lattice data; there is some deviation in the ultraviolet regime, although, as stated above, the logarithmic scaling is reproduced well. 

\begin{figure}[ht]
\includegraphics[width=10cm,angle=0]{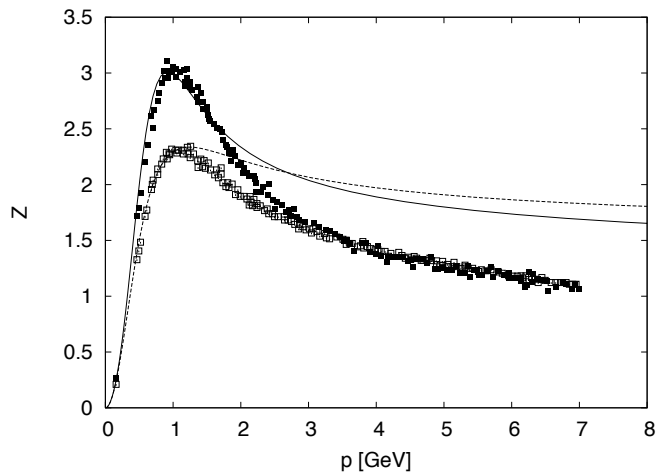}
\caption{Quenched data (solid points); quenched SDE (solid line); $n_f=3$ lattice (open points); unquenched SDE ($n_f=3$, dashed line). Data from \protect\cite{B2}.}
\label{fig-Z-comp}
\end{figure}

Fig. \ref{fig-gluon-lattice} shows that the gluon dressing function follows the lattice results closely in the infrared region. The gluonic propagator is often represented in terms of the dressing function $Z(p^2)$, which stresses the intermediate momentum region and has a distinct peak at $p \approx 1$ GeV. This is also accurately reproduced with our parameter set, as seen in Fig. \ref{fig-Z-comp}. Unfortunately, the ultraviolet scaling behavior of Eq. \ref{eq-g} is not recovered. One expects $\gamma(n_f=0) = -0.59$, while typical numerical results are $\gamma \approx -0.29$. The lattice data of Fig. \ref{fig-Z-comp} is obtained from Ref. \cite{B2}. This data has been scaled by a factor of 1.4 to bring it into agreement with the normalization of Ref. \cite{lattice-gluon}. Ref. \cite{B2} also presented the gluon propagator computed with two chiral and one light quark species (open squares in Fig. \ref{fig-Z-comp}). We have solved the SDE with three chiral quarks as a rough test of unquenching effects. The result, shown as a dashed line,  reproduces the infrared suppression seen in the lattice computation, although both curves do not scale correctly in the ultraviolet region. Both theory curves were generated with the renormalization condition $G(2.66\ {\rm GeV}) = 0.3$ GeV$^{-2}$.

\begin{figure}[ht]
\includegraphics[width=10cm,angle=0]{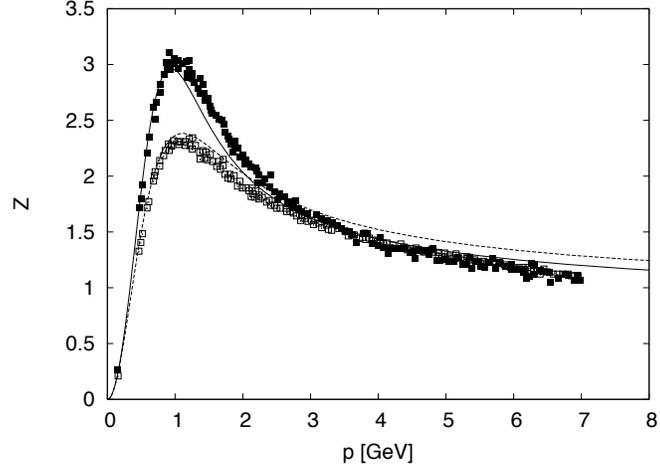}
\caption{Quenched data (solid points); quenched SDE (solid line); $n_f=3$ lattice (open points); unquenched SDE (dashed line). Data from \protect\cite{B2}. Simple three-gluon vertex model with $G(0.5\ {\rm GeV}, n_f=0)=7.5$ GeV$^{-2}$, $G(0.5\ {\rm GeV}, n_f=3) = 6.0$ GeV$^{-2}$, and $g_q=0.57$.}
\label{fig-Z-comp3}
\end{figure}

Since all contributions to the gluon propagator are included in the formalism and we have argued that the quark-gluon and ghost-gluon vertices are reasonably accurate, the missing ultraviolet strength of the polarization scalars must reside in the three- or four-gluon vertices. We have tested this by multiplying the tree-level three-gluon vertex by the factor $1 +\log(p_1^2 + p_2^2 + p_3^2)$, where the momenta are associated with each leg of the vertex. No parameterization of the vertex form was attempted for this simple exercise. Fig. \ref{fig-Z-comp3} shows that straightforward modifications such as this can yield a gluon propagator that is in decent agreement with lattice results.

\begin{figure}[ht]
\includegraphics[width=10cm,angle=0]{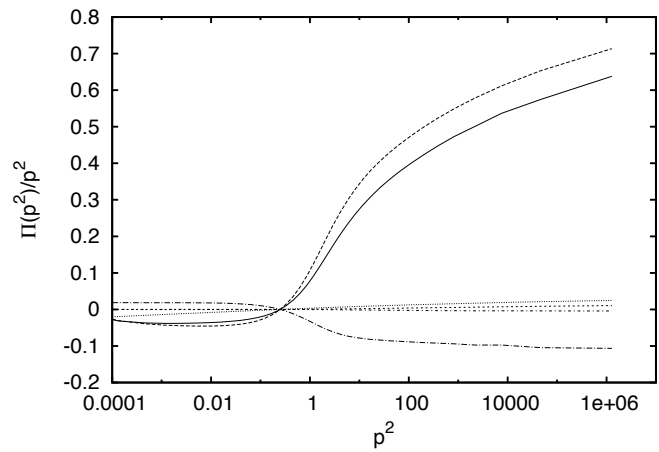}
\caption{Total (solid line), two-gluon (dashed), quark (small dash), ghost (dotted), sunset (dash-dot), and squint (long dash-dot) polarization functions.}
\label{fig-pi-comp3}
\end{figure}

Finally, the relative sizes of the various contributions to the gluonic dressing function are of interest. These are shown in Fig. \ref{fig-pi-comp3}. One sees that the gluonic loop dominates the right hand side of the gluonic gap equation. Alternatively, the quark, ghost, and sunset loop contributions are quite small and have negligible effect on the numerical value of the gluon dressing function. The squint vacuum polarization is about 15\% of the magnitude of the gluonic loop and of opposite sign. This has the effect of increasing $G$ in the ultraviolet region, which worsens the agreement with lattice results. There is also a slight degradation in agreement near $p= 0.5$ GeV.  The fact that the sunset diagram is much smaller than the squint agrees with the analysis of Bloch\cite{2loop}.

\section{Conclusions}

We have presented the first computation of the gluon propagator with the complete Schwinger-Dyson equations at the vertex model level. In particular, the two-loop squint and sunset diagrams have been included. A simple method to deal with quadratic divergences has been suggested and appears to produce satisfactory results. Specifically, we have found no problems with rendering the two-loop diagrams finite.

We have argued that it is important to work with renormalized quantities since this disambiguates vertex modelling from renormalization issues. 
This in turn forces recognition of the 
violation of gauge symmetry that is induced by truncation in the $n$-point functions. We suggest dealing with this issue by allowing different bare couplings for each interaction in the QCD Lagrangian. Presumably as the accuracy of the system of equations is improved, Slavnov-Taylor constraints will be better obeyed and the renormalized couplings will approach a common value.

We have tested the latter assertion by comparing a rainbow-ladder calculation with one with simple model vertices. The couplings move from 

\be
g_q = 3.03 \qquad g_{c} = 1.12 \qquad g_3 = 0.6
\ee
to
\be
g_q = 1.35 \qquad g_{c} = 1.12 \qquad g_3 = 0.6,
\ee
when the CBC vertex is used. The quark-gluon coupling drops further to $g_q=1.31$ when the simple three-gluon vertex model is used. Both effects are in agreement with our claim. 

We have found that the CBC vertex is required to obtain a stable  solution, dynamical chiral symmetry breaking, and the correct asymptotic behavior of the dynamical mass. Alternatively, the tree ghost-gluon vertex yields a ghost dressing function in  good agreement with lattice results, while using a model vertex can lead to instability in the iterative solution, depending on the couplings.
The gluon dressing function agrees well with a lattice computation except in the ultraviolet region. Our experience with the dynamical mass suggests that a good three-gluon vertex model will likely correct this problem.

We are able to compare the relative importance of various loop diagrams to the gluon propagator and find the hierarchy

\be
\Pi_{\rm gluon} \gg \Pi_{\rm squint} \gg \Pi_{\rm sunset} \approx \Pi_{\rm ghost} \approx \Pi_{\rm quark}.
\ee
This does not agree with the results of von Smekal {\it et al.} and other groups, who find that the ghost loop dominates the gluon propagator. We suspect that this is due to the choice of three-gluon vertex model. For example, many authors use a model that has inverse powers of the gluon dressing function that strongly attenuates the contribution of the gluon propagator to the gluon loop polarization scalar (in fact, some models eliminate it entirely). We have also not been able to reproduce the scaling solutions, where 
$G(k\to 0) = 0$. Again, this is likely due to model vertex choices.

Multiplicative renormalizability is often invoked while building vertex models. This is typically discussed in two ways: (i) vertex models are constructed in an attempt to mimic one-loop resummed perturbation theory or (ii) the restraints of the finite renormalization group are maintained with appropriate vertex choices. The usual application of the latter view is in requiring that the dynamical quark mass is a renormalization group invariant. It is possible to achieve this by using cutoff-dependent vertex models, but we have argued against this approach since it undermines renormalization. In fact, simply using the standard renormalization procedure of, e.g.,  Eqs. \ref{eq-S} and \ref{eq-G} in the quark gap equations guarantees that the dynamical mass is renormalization group invariant. This implies, once again, that model vertices should be renormalized. Renormalization group invariance of the dynamical quark mass implies that it is independent of the renormalization scale, which implies that it is a function of the running coupling only. We have confirmed this by computing the dynamical mass for a variety of renormalization scales with results that match those of Fig. \ref{fig-quark-lattice}.


As a technical point, we have found that large grids are required to compute the two-loop diagrams accurately and that differentiable dressing functions are needed for subtractions to yield finite results in two-loop diagrams.

With consistent renormalization in place and all diagrams taken into account we are able to state that a model with a tree ghost-gluon vertex is likely adequate and enhancement is required in the quark-gluon vertex to obtain dynamical chiral symmetry breaking and the correct asymptotic behavior of the dynamical quark mass. A similar enhancement is necessary in the three-gluon vertex, while little can be said about the four-gluon vertex at this point except that it likely does not affect the two-point functions strongly. Further work incorporating lattice or Schwinger-Dyson models of the three-gluon vertex will be of use. Extensions to finite temperature and other observables are also clearly of interest.

Finally, we remark that many Schwinger-Dyson approaches have been able to reproduce the main features of the nonperturbative propagators of QCD. Evidently the simple vertex models and truncations typically used are capable of capturing the main features of the physics driving these quantities. That this is true is something of a mystery since no small parameter has been exploited to guarantee the success of this approach. Perhaps one can conclude that two-point functions are relatively simple quantities. Nevertheless, the apparent successes at this level cannot be assumed to continue with higher $n$-point functions. For example, the diagrams that contribute to the quark-antiquark interaction can be topologically complicated and involve nonperturbative vertices at all orders.

\acknowledgments
This research was supported by the U.S. Department of Energy under contract
DE-FG02-00ER41135. We thank C. Fischer, R. Williams, and D. Wilson for discussions.

\end{document}